\documentclass[conference]{IEEEtran}
\IEEEoverridecommandlockouts
\usepackage{amsmath,amssymb,amstext}
\usepackage{graphicx}

\begin{document}
\title{On Capacity of Active Relaying in Magnetic Induction based Wireless Underground Sensor Networks}
\author{S. Kisseleff $^\blacktriangle$, B. Sackenreuter $^{\blacklozenge}$, I. F. Akyildiz $^{\bigstar}$, and W. Gerstacker $^\blacktriangle$\\
$^\blacktriangle$ Institute for Digital Communications, Friedrich-Alexander University (FAU) \\
Erlangen-N\"urnberg, Germany, \{kisseleff, gersta\}@lnt.de\\
$^{\blacklozenge}$ Fraunhofer Institute for Integrated Circuits, N\"urnberg, Germany, benjamin.sackenreuter@iis.fhg.de\\
$^{\bigstar}$ Broadband Wireless Networking Lab, Georgia Institute of Technology, USA, ian@ece.gatech.edu
\thanks{This work was supported by the German Research Foundation (Deutsche Forschungsgemeinschaft, DFG) under Grant No. GE 861/4-1}
}
\maketitle
\begin{abstract}
Wireless underground sensor networks (WUSNs) present a variety of new research challenges. Magnetic induction (MI) based transmission has been proposed to overcome the very harsh propagation conditions in underground communications in recent years. In this approach, induction coils are utilized as antennas in the sensor nodes. This solution
achieves  longer transmission ranges compared to the traditional electromagnetic (EM) waves based approach. Furthermore, a passive relaying technique has been proposed in the literature where additional resonant circuits are deployed between the nodes. However, this solution is shown to provide only a limited performance improvement under practical system design contraints. In this work, the potential of an active relay device is investigated which may improve the performance of the system by combining the benefits of the traditional wireless relaying and the MI based signal transmission.
\end{abstract}
\section{Introduction}
\label{sec:1}
The objective of Wireless Underground Sensor Networks (WUSNs) is to establish an efficient wireless communication in the underground medium. Typical applications for such networks include soil condition monitoring, earthquake prediction, border patrol, etc. \cite{WSN_survey}, \cite{WUSN_reschall}. Since the propagation medium is
soil, rock, and sand, traditional wireless signal propagation techniques using electromagnetic (EM) waves can be only applied for very small transmission ranges due to a high
path loss and vulnerability to changes of soil properties, such as moisture or water contents \cite{Sig_propag_underground}.
\newline
Magnetic induction (MI)-based WUSNs were first introduced in \cite{WUSN_reschall}, \cite{MI_comms_WUSN} and make use of magnetic antennas implemented as coils. In recent years, some system models have been developed in order to investigate the system behavior and typical properties of the MI based transmission channels,  e.g. \cite{MI_comms_WUSN}, \cite{cap_perf_near_field}, and \cite{power_and_cap_MI}. More realistic channel and noise models for a point-to-point transmission were derived in \cite{pract_ch_cap}. These models incorporate the losses due to the transmission medium and the power reflections between the coils. Furthermore, it was shown that the transmission through a conductive medium like soil can only be established using a carefully optimized set of system parameters. Recently, novel modulation approaches for MI based transmissions using practical signal processing components (transmit, receive, and equalization filters) have been proposed in \cite{modul_MI}.\\
The idea of deploying additional resonance circuits (passive relay devices) between two transceiver nodes and combining them into waveguide structures has been first proposed in \cite{MI_waveguide_first} for MI based communications systems and in \cite{MI_comms_WUSN} for MI-WUSNs. Similar to traditional wireless relaying concepts this approach is supposed to benefit from a lower path loss. However, as it was pointed out in \cite{pract_ch_cap}, the MI waveguides show limited performance in the underground medium under realistic design constraints due to a very narrow bandwidth, as discussed in detail in \cite{modul_MI}. Hence, in this work, we propose to add a further functionality to the relay devices, such that the received signal at the relay is actively processed and retransmitted. This additional functionality for MI enabled communication systems is already  mentioned in \cite{agbinya_masihpour_relaying}. However, the potential of this technique for the MI-WUSNs has not been shown and analyzed as of to date. Hence, we investigate the properties of active relaying schemes in MI-WUSNs. Furthermore, we consider the possibility of establishing a full duplex relaying mode, which may dramatically improve the performance of this scheme.\\
This paper is organized as follows. In Section \ref{sec:2}, the system model is derived, which incorporates the influence of a single active relay onto the signal transmission and reception. To this end, we investigate the frequency-selective path loss of the relayed transmission and the colored noise at the receiver front-end. In Section \ref{sec:3} the problem of system parameter selection and power allocation is considered. Numerical results are provided in Section \ref{sec:4} and Section \ref{sec:5} concludes the paper.
\section{System Model}
\label{sec:2}
\begin{figure}
\centering
\includegraphics[width=0.48\textwidth]{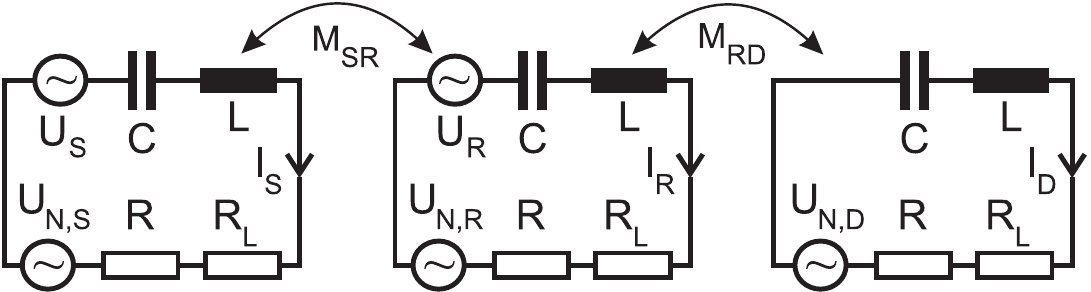}
\caption{Equivalent circuit model for the source-relay-destination link.}
\label{circuits}
\end{figure}

In this work, we derive new channel and noise models for direct MI transmission in presence of a single active relay. We assume a simple MI relay network with two MI transceivers (called source and destination) and one relay between them. All three devices contain the same set of passive elements of the resonance circuit: 
\begin{itemize}
\item an induction coil with inductivity $L$, which is modeled as a multilayer air core coil,
\item a resistor with resistance $R$, which models the parasitic wire resistance,
\item a capacitor with capacitance $C$, which is tuned to make the circuit resonant at the carrier frequency $f_0=\frac{1}{2\pi\sqrt{LC}}$,
\item a load resistor $R_L$, which is optimized for minimum power reflections at the transceivers for transmissions at resonance frequency.
\end{itemize}
The basic structure of the relay network is shown in Fig. \ref{circuits}.
The coupling between the source and the relay is determined by the mutual inductance $M_{SR}$, and between the relay and the destination by the mutual inductance $M_{RD}$, respectively. These mutual inductances are calculated similarly to \cite{pract_ch_cap},
\begin{eqnarray}
\label{eq:2_1}
\vspace*{-2mm}
M_{SR}\hspace*{-2mm}&=&\hspace*{-2mm}\mu\pi N^2\frac{a^4}{4r_{SR}^3}  J_{SR}  G_{SR},\\
M_{RD}\hspace*{-2mm}&=&\hspace*{-2mm}\mu\pi N^2\frac{a^4}{4r_{RD}^3}  J_{RD}  G_{RD},
\vspace*{-2mm}
\end{eqnarray}
where $r_{\{ \}}$ denotes the distance between the specified two coils, $a$ stands for the coil radius, $N$ is the number of windings, and $\mu$ denotes the permeability of the medium. $J_{\{ \}}$ represents the polarization factor according to \cite{Interference_polariz}. $G_{\{ \}}$ is an additional loss factor due to eddy currents as pointed out in \cite{pract_ch_cap}. We assume that all devices are deployed in a homogeneous conductive environment (soil) with constant properties over space and time.\\
Similarly to the previous work (cf. \cite{pract_ch_cap}), we obtain the transmit power spectral density at the source node
\vspace*{-1mm}
\begin{equation}
\label{eq:2_2}
P_{St}(f)=\frac{\left|U_{S}\right|^2}{2}\left|\frac{Z_g^2+(2\pi fM_{RD})^2}{Z_g(Z_g^2+(2\pi f)^2(M_{SR}^2+M_{RD}^2))}\right|,
\vspace*{-1mm}
\end{equation}
where $Z_g=j2\pi fL+\frac{1}{j2\pi fC}+R+R_L$ denotes the sum of all circuit impedances and $U_S$ is the input voltage at the source. Correspondingly, $U_R$ stands for the input voltage at the relay. \\
The receive power spectral density at the relay is given by
\vspace*{-1mm}
\begin{equation}
\label{eq:2_3}
P_{Rr}(f)=\frac{\left|U_{S}\right|^2}{2}\left|\frac{2\pi fM_{SR}}{Z_g^2+(2\pi f)^2(M_{SR}^2+M_{RD}^2)}\right|^2\hspace*{-1mm}R_L.
\vspace*{-1mm}
\end{equation}
Hence, the channel power gain can be calculated as $\left|H_{SR}(f)\right|^2=\frac{P_{Rr}(f)}{P_{St}(f)}$ using \eqref{eq:2_2} and \eqref{eq:2_3}. In addition, the transmitted signal from the source is received at the destination via passive relaying, which has been investigated in the context of MI waveguides in the previous works. The corresponding receive power spectral density is given by
\begin{equation}
\label{eq:2_4}
\hspace*{-2mm}P_{Dr1}(f)=\frac{\left|U_{S}\right|^2}{2}\left|\frac{(2\pi f)^2(M_{SR}M_{RD})}{Z_g(Z_g^2+(2\pi f)^2(M_{SR}^2+M_{RD}^2))}\right|^2\hspace*{-1mm}R_L,
\end{equation}
such that the channel power gain from source to destination can be given by $\left|H_{SD, \: \mathrm{p}}(f)\right|^2=\frac{P_{Dr1}(f)}{P_{St}(f)}$. When the relay transmits its signal, we obtain for the transmit power spectral density at the relay and for the receive power spectral density at the destination, respectively,
\vspace*{-1mm}
\begin{eqnarray}
\label{eq:2_5}
\hspace*{-2mm}P_{Rt}(f)\hspace*{-2mm}&=&\hspace*{-2mm}\frac{\left|U_{R}\right|^2}{2}\left|\frac{Z_g}{Z_g^2+(2\pi f)^2(M_{SR}^2+M_{RD}^2)}\right|,\\
\label{eq:2_6}
\hspace*{-2mm}P_{Dr2}(f)\hspace*{-2mm}&=&\hspace*{-2mm}\frac{\left|U_{R}\right|^2}{2}\left|\frac{j2\pi fM_{RD}}{Z_g^2+(2\pi f)^2(M_{SR}^2+M_{RD}^2)}\right|^2\hspace*{-1mm}R_L.
\end{eqnarray}
This yields a relay to destination channel power gain $\left|H_{RD}(f)\right|^2=\frac{P_{Dr2}(f)}{P_{Rt}(f)}$. 
A major difference compared to the relaying in traditional communication systems is an additional path loss, which basically results from the linear mapping of the received signal at the load impedance onto the transmit signal $U_R$. This mapping can be described using an additional channel gain
\hspace*{-2mm}
\begin{equation}
\label{eq:2_7}
\left|H_{RR}(f)\right|^2=\left|\frac{Z_g}{Z_g^2+(2\pi f)^2(M_{SR}^2+M_{RD}^2)}\right|R_L,
\end{equation}
such that for nonregenerative relaying schemes (e.g. amplify-and-forward relaying) the total channel power gain of the active relay link (without the amplification $A$ of the relay) can be given by
\begin{equation}
\label{eq:2_8}
\left|H_{SD, \: \mathrm{a}}(f)\right|^2=\left|H_{SR}(f)\right|^2\left|H_{RR}(f)\right|^2\left|H_{RD}(f)\right|^2.
\end{equation}
For the noise modeling, we focus on the thermal noise produced by the resistors in all three resonance circuits (voltage sources $U_{N,S}$, $U_{N,R}$ and $U_{N,D}$ in Fig. \ref{circuits}). The noise power spectral density at the relay and destination, respectively, can be calculated similarly to \cite{pract_ch_cap}
\begin{equation}
\label{eq:2_9}
P_{Rn}(f)\hspace*{-0.5mm}=\hspace*{-0.5mm}4KT(R_LR+R_L^2)\frac{\left|Z_g\right|^2\hspace*{-0.5mm}+\hspace*{-0.5mm}(2\pi f)^2(M_{SR}^2+M_{RD}^2)}{2\left|Z_g^2\hspace*{-0.5mm}+\hspace*{-0.5mm}(2\pi f)^2(M_{SR}^2+M_{RD}^2)\right|^2},
\end{equation}
\begin{eqnarray}
\label{eq:2_10}
\hspace*{-4mm}P_{Dn}(f)\hspace*{-2mm}&=&\hspace*{-2mm}\frac{4KT(R_LR+R_L^2)\left|Z_g^2-(2\pi fM_{SR})^2\right|^2}{2\left|Z_g(Z_g^2+(2\pi f)^2(M_{SR}^2+M_{RD}^2))\right|^2}\notag\\
\hspace*{-4mm}\hspace*{-2mm}&&\hspace*{-2mm}+\hspace*{1mm}\frac{4KT(R_LR+R_L^2)(2\pi f)^4(M_{SR}M_{RD})^2}{2\left|Z_g(Z_g^2+(2\pi f)^2(M_{SR}^2+M_{RD}^2))\right|^2}\notag\\
\hspace*{-4mm}\hspace*{-2mm}&&\hspace*{-2mm}+\hspace*{1mm}\frac{4KT(R_LR+R_L^2)\left|Z_g(2\pi fM_{RD})\right|^2}{2\left|Z_g(Z_g^2+(2\pi f)^2(M_{SR}^2+M_{RD}^2))\right|^2},
\end{eqnarray}
where $K\approx1.38\cdot 10^{-23}$ J/K is the Boltzmann constant and $T$ is the temperature in Kelvin ($T=290$ K in this work).\\
Since our main focus is on the information theoretic analysis of the proposed scheme, the performance metrics are related to the achievable data rates, which are calculated using Shannon's channel capacity formula.
\section{Active Relaying}
\label{sec:3}
In this section we investigate the properties of different active relaying schemes such as amplify-and-forward (AF), frequency-selective amplify-and-forward (also called filter-and-forward, FF) or decode-and-forward (DF) known from the literature (e.g. \cite{cooperation_book1}, \cite{cooperation_book2}). In the commonly used half duplex relaying mode \cite{cooperation_book1}, the transmission and reception data streams are separated e.g. by means of time-division duplex (TDD). However, the corresponding loss in data rate can be reduced by applying a full duplex approach. For simplicity, we restrict ourselves to the half duplex (HD) relaying mode in Section \ref{sec:3_1} - Section \ref{sec:3_3}. In Section \ref{sec:3_4}, the modifications of the system are discussed, which are needed in order to enable the full duplex (FD) relaying mode.\\
There are several system parameters, which can be optimized for any relaying scheme. Some parameters are typical for the MI based transmissions, as discussed in \cite{pract_ch_cap}, e.g., the resonance frequency $f_0$ and the number of coil windings $N$. Other parameters are typical for the relayed transmission, e.g. the transmit power at the relay $P_R$ or the relay position relatively to the transceiver nodes. All these parameters (except for the design of the transmit filters) are optimized via a full search similarly to \cite{pract_ch_cap}. The optimization of the transmit filters for different relaying schemes is discussed below.
\subsection{Amplify-and-forward relaying (AF)}
\label{sec:3_1}
\begin{figure}
\centering
\includegraphics[width=0.48\textwidth]{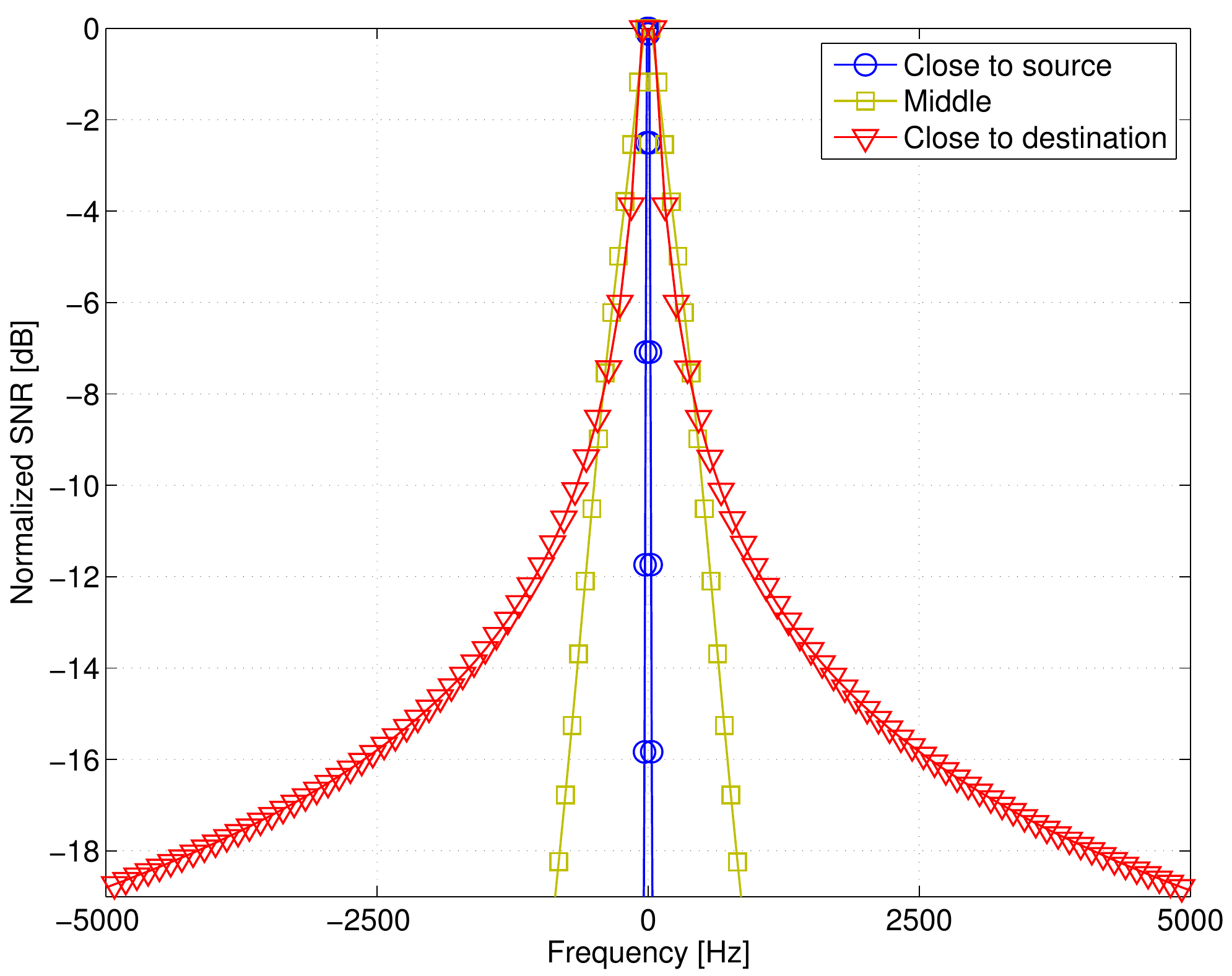}
\caption{Normalized $\mathrm{SNR}_{MRC}(f)$ of the received baseband signal at the Destination using AF relaying for different positions of the Relay.}
\label{af_snr}
\end{figure}

The simplest relaying scheme is AF relaying. Here, the signal received in the first time slot at the relay is amplified using a frequency flat constant $A$ and sent to the destination in the second time slot. $A$ is related to the available relay transmit power $P_R$ and can be given by
\begin{equation}
\label{eq:3_1_1}
A=\sqrt{\frac{P_R}{\displaystyle\int_B\left(P_{St}(f)\left|H_{SR}(f)\right|^2+P_{Rn}(f)\right)\left|H_{RR}(f)\right|^2\mathrm{d}f}},
\end{equation}
where $B$ represents the chosen signal bandwidth. The total received power spectral density at the destination in the second time slot is
\begin{eqnarray}
\label{eq:3_1_2}
\hspace*{-3mm}P_{Dr2}(f)\hspace*{-2mm}&=&\hspace*{-2mm}A^2\left(P_{St}(f)\left|H_{SR}(f)\right|^2+P_{Rn}(f)\right)\\
\hspace*{-3mm}\hspace*{-2mm}&&\hspace*{-2mm}\times\hspace*{1mm}\left|H_{RR}(f)\right|^2\left|H_{RD}(f)\right|^2+P_{Dn}(f)\notag\\
\hspace*{-3mm}\hspace*{-2mm}&=&\hspace*{-2mm}A^2P_{St}(f)\left|H_{SD, \: \mathrm{a}}(f)\right|^2\notag\\
\hspace*{-3mm}\hspace*{-2mm}&&\hspace*{-2mm}+\hspace*{1mm}A^2\left|H_{RR}(f)\right|^2\left|H_{RD}(f)\right|^2\hspace{-0.5mm}P_{Rn}(f)\hspace{-0.5mm}+\hspace{-0.5mm}P_{Dn}(f).\notag
\end{eqnarray}
In order to improve the performance of this scheme, we additionally employ the signal received from the source at the destination in the first time slot via passive relaying, which has a power spectral density
\begin{equation}
\label{eq:3_1_3}
P_{Dr1}(f)=P_{St}(f) \left|H_{SD, \: \mathrm{p}}(f)\right|^2+P_{Dn}(f).
\end{equation}
The received signals from two time slots may be affected by different channel transfer function. Therefore, we combine them coherently using the optimal approach of maximum ratio combining (MRC). For this, we employ the respective matched filters at the destination. The data rate of the relay network can be given by
\begin{equation}
\label{eq:3_1_4}
C_{AF, HD}=\frac{1}{2}\int_{B}{\log_2\left(1+\mathrm{SNR}_{MRC}(f)\right)\mathrm{d}f},
\end{equation}
where we define $\mathrm{SNR}_{MRC}(f)=\mathrm{SNR}_1(f)+\mathrm{SNR}_2(f)$ with 
\begin{eqnarray}
\label{eq:3_1_41}
\hspace*{-5.5mm}\mathrm{SNR}_1(f)\hspace{-3mm}&=&\hspace{-3mm}\frac{P_{St}(f)\left|H_{SD, \: \mathrm{p}}(f)\right|^2}{P_{Dn}(f)},\\
\label{eq:3_1_42}
\hspace*{-5.5mm}\mathrm{SNR}_2(f)\hspace{-3mm}&=&\hspace{-3mm}\frac{A^2P_{St}(f)\left|H_{SD, \: \mathrm{a}}(f)\right|^2}{A^2\left|H_{RR}(f)\right|^2\hspace*{-0.5mm}\left|H_{RD}(f)\right|^2\hspace*{-0.5mm}P_{Rn}(f)\hspace*{-0.5mm}+\hspace*{-0.5mm}P_{Dn}(f)}.
\end{eqnarray} 
Since $P_{St}(f)$ can only be optimized for a fixed value of $A$ and $A$ depends on $P_{St}(f)$ according to \eqref{eq:3_1_1}, we propose an iterative algorithm in order to maximize the achievable data rate. Starting with a uniform power distribution, the value of $A$ is calculated in each iteration based on \eqref{eq:3_1_1}. Then, the optimal power allocation $P_{St}(f)$ is calculated using the waterfilling approach. Since $\mathrm{SNR}_{MRC}(f)$ has a shape $P_{St}(f)  K(f)$, where $K(f)$ is independent of $P_{St}(f)$, the waterfilling solution can be given by
\vspace*{-1mm}
\begin{equation}
\label{eq:3_1_5}
P_{St}(f)=\max\lbrace\frac{1}{\lambda}-\frac{1}{K(f)}, \ 0\rbrace,
\vspace*{-1mm}
\end{equation}
where $\frac{1}{\lambda}$ stands for the water level and is calculated according to the power constraint. \\
As opposed to traditional wireless communication systems, AF relaying is not well suited for MI based transmissions, due to a high frequency-selectivity of the channel, especially if the relay is placed close to the source. This yields a narrow frequency band and a poor system performance. 
This behavior can be explained via Fig. \ref{af_snr}, where normalized results for $\mathrm{SNR}_{MRC}(f)$ are shown using the optimal system parameters for the corresponding relay positions.
If we neglect the passive relaying link due to much higher path loss, we can distinguish between three cases:
\begin{enumerate}
\item Relay is placed close to the source: $P_{Dn}(f)$ is the dominant noise source, such that 
\vspace*{-1mm}
\begin{equation}
\mathrm{SNR}_2(f)=\frac{A^2P_{St}(f)\left|H_{SD, \: \mathrm{a}}(f)\right|^2}{P_{Dn}(f)}
\end{equation}
results. Due to an extremely frequency-selective path loss $\left|H_{SD, \: \mathrm{a}}(f)\right|^2$, the useful bandwidth is a few Hz.
\item Relay is placed close to the destination: $A^2\left|H_{RR}(f)\right|^2\left|H_{RD}(f)\right|^2P_{Rn}(f)$ is the dominant noise source, such that
\vspace*{-2mm}
\begin{equation}
\mathrm{SNR}_2(f)=\frac{P_{St}(f)\left|H_{SR}(f)\right|^2}{P_{Rn}(f)}
\end{equation}
results. Here, the frequency-selectivity is identical with that of the direct MI transmission, which provides the largest possible bandwidth of up to several 10 kHz.
\item Relay is close to the middle: no noise component can be viewed as clearly dominant. The resulting bandwidth is larger than for case 1, but smaller than for case 2.
\end{enumerate}
Hence, the achievable data rate for AF relaying is maximum at the destination and the performance of this scheme is in general very poor. 
\subsection{Filter-and-Forward relaying (FF)}
\label{sec:3_2}
The FF relaying is similar to the AF relaying, however, a frequency-selective amplification coefficient $A(f)$ is used instead of a constant $A$. As known from the literature (e.g. \cite{ff_optimize}, \cite{ff_mimo}), $A(f)$ can be optimized in order to maximize the achievable data rate in wireless communication systems. Hence, the optimization problem can be formulated as follows:
\vspace*{-3mm}
\begin{eqnarray}
\label{eq:3_2_1}
&&\hspace*{-14mm}\max_{A(f), P_{St}(f)\in \mathbb{R}_0^+}\int_{B}{\log_2\left(1+\mathrm{SNR}_{MRC}(f)\right)\mathrm{d}f},\\
\mbox{s.t.:}
&1)& P_S=\int_BP_{St}(f)\mathrm{d}f, \notag\\
&2)& P_R=\int_BP_{Rt}(f)\mathrm{d}f.\notag
\end{eqnarray}
Unfortunately, it can be shown that this problem is not convex and its solution cannot be calculated using the well-known methods of convex optimization \cite{bconvex}. Hence, an iterative algorithm is proposed in \cite{ff_optimize}. In each iteration, two subproblems of \eqref{eq:3_2_1} need to be solved. By solving the first subproblem, $P_{St}(f)$ for a given relay transmit power spectrum density $P_{Rt}(f)$ is optimized. Then, by solving the second subproblem, $P_{Rt}(f)$ for a given power distribution $P_{St}(f)$ is optimized. $P_{Rt}(f)$ is related to the amplification $A(f)$ by 
\vspace*{-1.5mm}
\begin{equation}
\label{eq:3_2_2}
P_{Rt}(f)=\left|A(f)\right|^2\left(P_{St}(f)\left|H_{SR}(f)\right|^2+P_{Rn}(f)\right)\left|H_{RR}(f)\right|^2.
\end{equation}
According to \cite{ff_optimize}, both subproblems are convex, such that a closed-form solution for each subproblem can be found using a Lagrangian approach which is also given in \cite{ff_optimize}.
\subsection{Decode-and-forward relaying (DF)}
\label{sec:3_3}
In DF relaying, there is no linear mapping between the received signal and the transmitted signal at the relay, such that the two links (source $\rightarrow$ relay) and (relay $\rightarrow$ destination) can be decoupled and optimized separately. Due to a much higher attenuation of the passive relaying link, no diversity combining is needed at the destination node. The achievable data rate of the entire system can be given by
\begin{eqnarray}
\label{eq:3_3_1}
\hspace*{-6mm}C_{DF, HD}\hspace*{-2mm}&=&\hspace*{-2mm}\frac{1}{2}\min\{C_{SR}, C_{RD}\},\\
\label{eq:3_3_2}
\hspace*{-6mm}C_{SR}\hspace*{-2mm}&=&\hspace*{-2mm}\int_B{\log_2\left(1+\frac{P_{St}(f)\left|H_{SR}(f)\right|^2}{P_{Rn}(f)}\right)\mathrm{d}f},\\
\label{eq:3_3_3}
\hspace*{-6mm}C_{RD}\hspace*{-2mm}&=&\hspace*{-2mm}\int_B{\log_2\left(1+\frac{P_{Rt}(f)\left|H_{RD}(f)\right|^2}{P_{Dn}(f)}\right)\mathrm{d}f},
\end{eqnarray} 
The data rate \eqref{eq:3_3_1} is then maximized using the waterfilling rule applied to $P_{St}(f)$ and $P_{Rt}(f)$, respectively.
\subsection{Full duplex relaying mode (FD)}
\label{sec:3_4}
In contrast to the traditional communication systems, the use of full duplex relaying mode is possible with MI based transmissions due to the mostly time-invariant transmission channels, especially in MI based WUSNs as mentioned in \cite{chest_MI_2014}. For this, the known signal transmitted by the relay is subtracted from the sum of transmitted and received signals at the load impedance $R_L$ of the relay. The received own signal at the load impedance of the relay during transmissions from the relay can be given by
\vspace*{-1mm}
\begin{equation}
U_{Rr}=U_R \frac{Z_gR_L}{Z_g^2+(2\pi f)^2(M_{SR}^2+M_{RD}^2)},
\end{equation}
such that the corresponding path loss is equivalent to $\left|H_{RR}(f)\right|^2$ from \eqref{eq:2_7}. Since $U_R$ and all system parameters are known to the transmitting relay, in principle it can perfectly separate the own transmitted stream from the other received signals. Of course, in case of sudden changes of the propagation characteristics (due to rainfall, etc.) $M_{SR}$ and $M_{RD}$ may change, such that extraction of the desired signal becomes inefficient. However, this problem can be handled by a proper channel estimation, which is very efficient in WUSNs due to the stationary deployment. Hence, in the following we assume that a perfect signal extraction in full duplex mode is possible.\\
Since the received signals in the same time slot from the relay and the source (via passive relaying) cannot be combined in case of FD mode, these signals impose additional interference to each other. We focus on the stronger signal, which is obviously the signal coming from the relay to the destination. The second signal coming passively from the source is considered as interference. Hence, some equations shown in this section need to be changed accordingly:
\subsubsection{AF and FF relaying}
\label{sec:3_4_1}
For the AF and FF relaying schemes, we obtain
\vspace*{-1mm}
\begin{equation}
\label{eq:3_4_1_1}
\mathrm{SINR}_{FD}(f)=\frac{\mathrm{SNR}_2(f)}{1+\mathrm{SNR}_1(f)\hspace*{-1mm}\left(1+\frac{\left|A(f)\right|^2\left|H_{SD, \: \mathrm{a}}(f)\right|^2P_{Rn}(f)}{\left|H_{SR}(f)\right|^2P_{Dn}(f)}\right)},
\end{equation}
where $\mathrm{SINR}_{FD}(f)$ denotes the total signal-to-interference-plus-noise ratio in full duplex mode at frequency $f$ as received by the destination. Here, the amplification coefficient $A(f)$ is either frequency flat (AF relaying) or frequency-selective (FF relaying). With this, the achievable data rate can be given by
\begin{equation}
\label{eq:3_4_1_2}
C_{\{AF, FF\}, FD}=\int_B\log_2\left(1+\mathrm{SINR}_{FD}(f)\right)\mathrm{d}f.
\end{equation}
\subsubsection{DF relaying}
\label{sec:3_4_2}
For the DF relaying, only $C_{RD}$ needs to be adjusted, since $C_{SR}$ is not effected by the switch to the full duplex mode (assuming perfect extraction of the received signal is possible at the relay). Hence, we modify \eqref{eq:3_3_3} similarly to \eqref{eq:3_4_1_1} and \eqref{eq:3_4_1_2}:
\vspace*{-0.5mm}
\begin{equation}
\label{eq:3_4_2_1}
C_{RD, FD}\hspace*{-0.5mm}=\hspace*{-1mm}\int_B\hspace*{-0.5mm}\log_2\left(\hspace*{-0.5mm}1+\frac{P_{Rt}(f)\left|H_{RD}(f)\right|^2}{P_{Dn}(f)+P_{St}(f)\left|H_{SD, \: \mathrm{p}}(f)\right|^2}\right)\hspace*{-0.5mm}\mathrm{d}f.
\end{equation}
Here, the signal transmission requires only one time slot. Hence, $C_{DF, FD}=\min\{C_{SR}, C_{RD, FD}\}$ holds.\\
\vspace*{-1mm}

In general, the optimization algorithms shown in the previous sections are valid for the full duplex mode as well. This is due to a very high path loss of the passive relaying link, such that the optimal system parameters and transmit filters do not change much compared to the half duplex mode. Therefore, we apply \eqref{eq:3_4_1_1}-\eqref{eq:3_4_1_2} only for the calculation of the final results after finding the optimal parameters $P_{St}(f)$ and $A$ for the AF and FF relaying. For DF relaying, \eqref{eq:3_4_2_1} replaces \eqref{eq:3_3_3} after the optimization of both $P_{St}(f)$ and $P_{Rt}(f)$ as described in Section \ref{sec:3_3}.
\section{Numerical Results}
\label{sec:4}
\begin{figure}
\centering
\hspace*{-1.5mm}\includegraphics[width=0.56\textwidth]{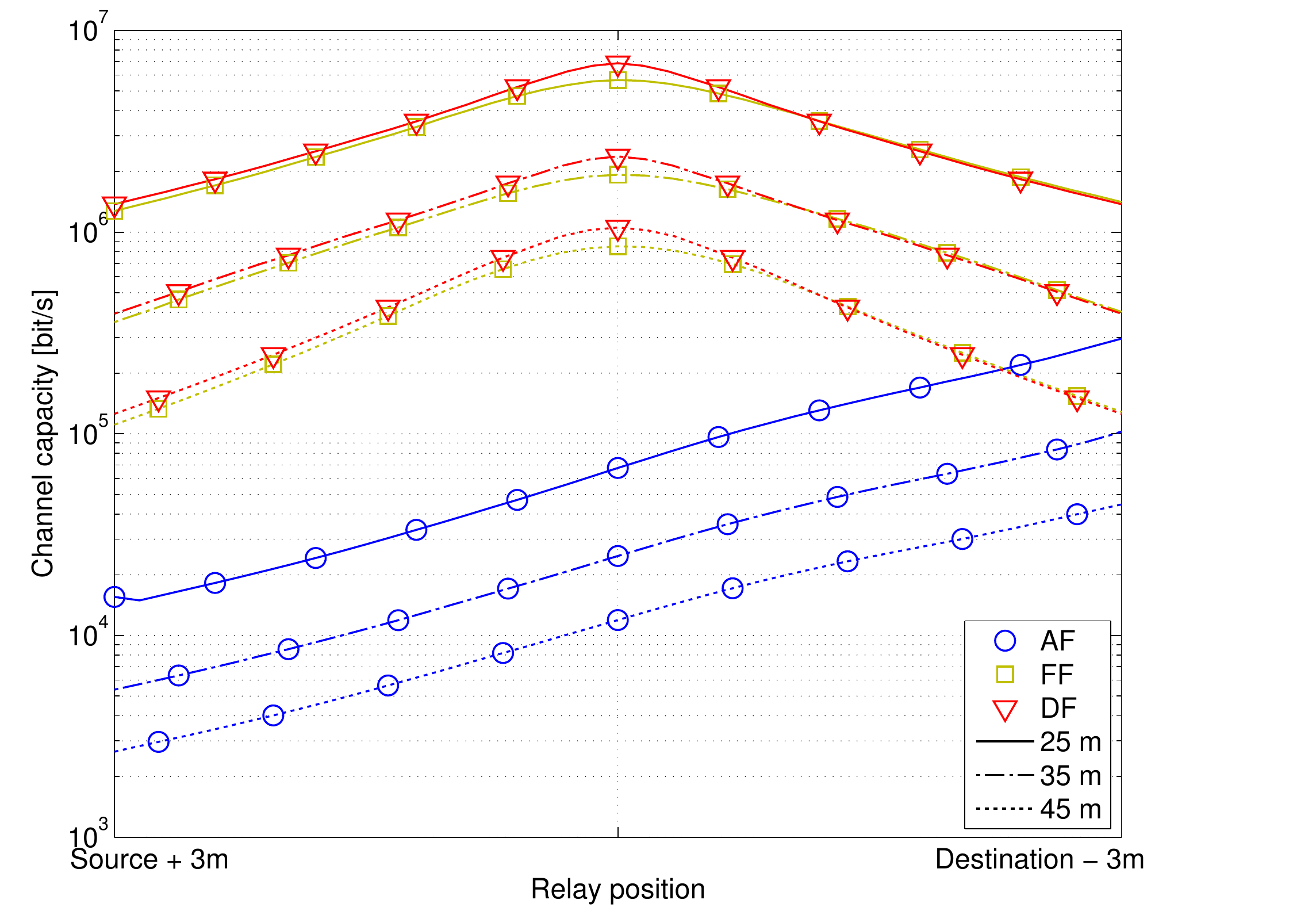}
\vspace*{-5mm}
\caption{Performance for different relay positions.}
\label{relpos}
\vspace*{-1mm}
\end{figure}

\begin{figure}
\centering
\includegraphics[width=0.48\textwidth]{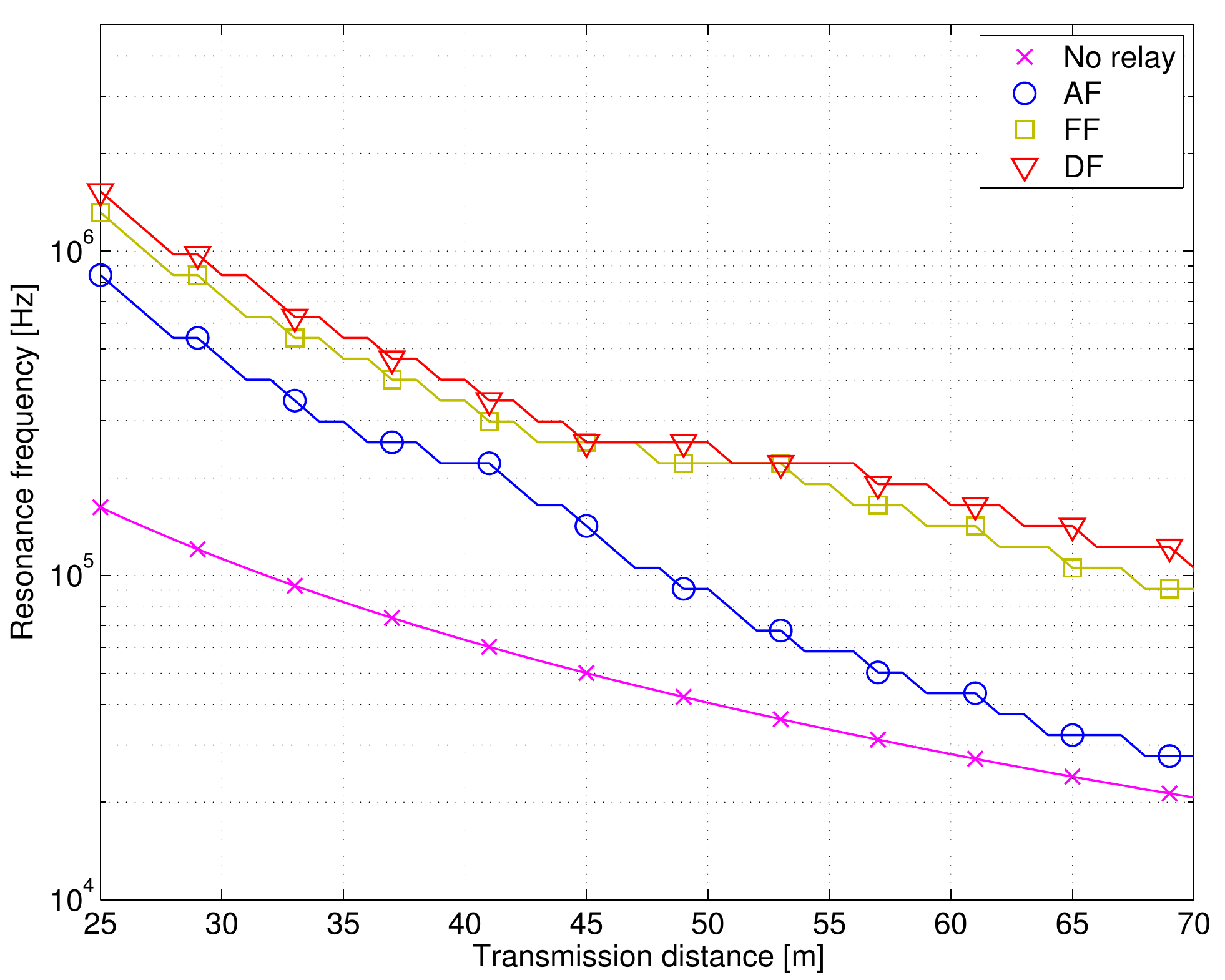}
\vspace*{-1mm}
\caption{Optimal resonance frequency.}
\label{optfreq}
\vspace*{-3mm}
\end{figure}

\begin{figure}
\centering
\includegraphics[width=0.48\textwidth]{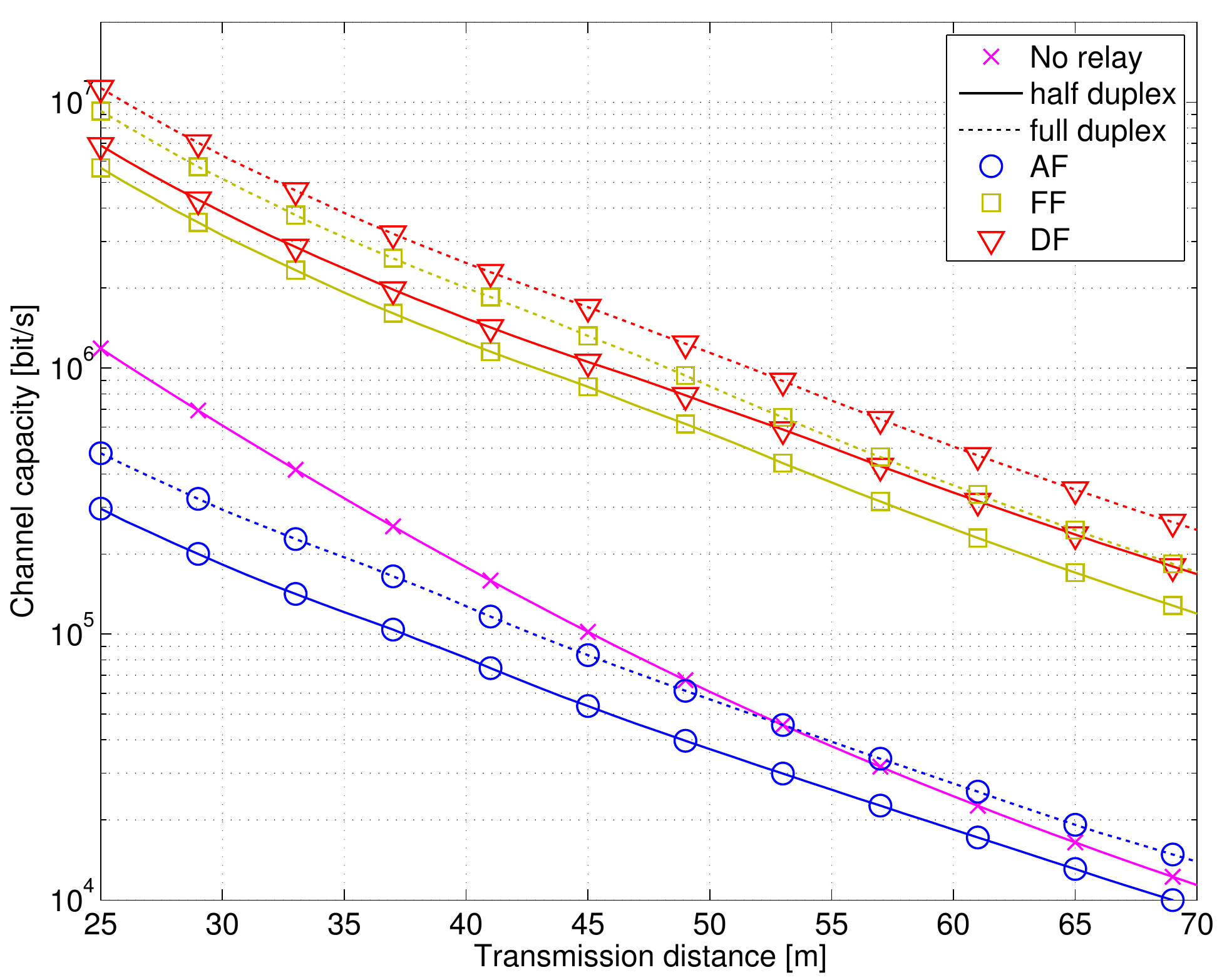}
\vspace*{-1mm}
\caption{Achievable data rate at the optimal relay position.}
\label{datarate}
\vspace*{-3mm}
\end{figure}

In this section, we discuss the performance of different relaying schemes based on the results obtained from a numerical evaluation of the proposed scheme. In our simulations, we assume a total transmit power budget of $P_{\mathrm{total}}$ = 10 mW. This includes the transmit power $P_S$ at the source and $P_R$ at the relay. The optimal values for $P_S$ and $P_R$ are found via full search under the constraint $P_S+P_R=P_{\mathrm{total}}$.
We utilize coils with wire radius 0.5 mm and coil radius $a$ = 0.15 m.  The maximum number of coil windings $N$ is 1000. The conductivity and permittivity of soil are $\sigma$ = 0.01 S/m and $\epsilon$ = 7$\epsilon_0$ for dry soil, respectively, where $\epsilon_0\approx 8.854\cdot 10^{-12}$ F/m. Since the permeability of soil is
close to that of air, we use $\mu$ = $\mu_0$ with the magnetic constant $\mu_0$ = 4$\pi\cdot 10^{-7}$ H/m. For a reduced path loss, we assume that the axes of all coils are parallel to the direction of the signal transmission, thus, the polarization factors $J_{SR}$ and $J_{RD}$ from \eqref{eq:2_1} are equal to two, which holds for horizontal axes deployment \cite{pract_ch_cap}. In order to investigate reasonable positions of the relay, we choose the shortest distance between the relay and any other transceiver to be at least 3 m.\\
At first, we show results for the achievable data rate using different relaying schemes in half duplex mode at different relay positions, see Fig. \ref{relpos}.
Here, the performance for AF relaying differs from the conventional wireless relaying due to its maximum close to the destination. However, even placing the AF relay close to the destination does not provide a sufficient data rate to justify the use and deployment of this device, as we shall see later. The results for FF and DF relaying correspond to our expectation, where the best performance is obtained if the relay is placed exactly in the middle.\\
Furthermore, we show the optimal resonance frequency $f_0$ for different transmission distances and different relaying schemes in half duplex mode, see Fig. \ref{optfreq}.
For comparison, we also show the results for the direct MI transmission (without relay). As known from the previous works on magnetic induction based systems (e.g. \cite{pract_ch_cap} or \cite{Interference_polariz}), the optimal resonance frequency decreases with increasing transmission distance. This behavior is confirmed here for the relay networks as well. Since the relay is placed in between the source and destination, the transmission distance of the longer link is smaller than the total transmission distance. Therefore, the optimal resonance frequency for all relaying schemes is higher than for the direct transmission. From the same reason, the resonance frequency for AF is lower than for FF and DF, which is due to the relay position in AF close to the destination. Moreover, the optimal resonance frequency for AF relaying converges to that of the direct MI transmission for increasing distance. For the relayed signal transmission, the curves obtained for the resonance frequency are not smooth over transmission distance, which is due to the finite precision of the optimization, where the resonance frequency is one of the many parameters jointly optimized via full search.\\
The achievable data rate is shown versus transmission distance in Fig. \ref{datarate}. 
Obviously, AF relaying does not improve the performance of an MI based transmission, as mentioned earlier. Hence, this type of relaying should not be used in MI-WUSNs. On the other hand, we observe a very good performance of the FF relaying scheme, which almost reaches the upper bound given by the DF relaying. Hence, even non-regenerative relaying strategies can be useful for MI-WUSNs. However, DF is still 20$\%$ - 40$\%$ better than FF, which is due to the noise enhancement at the relay according to \eqref{eq:3_2_2} for FF. However, DF requires a much higher complexity of the signal processing and transceiver design, which is an important criterion for the design of MI-WUSNs. \\
For full duplex relaying, the data rate is 50$\%$ - 65$\%$ larger than for the half duplex relaying in all considered relaying schemes, which makes this new technique very promising. Using DF in full duplex mode, data rate can be even increased by factor 10 - 22 compared to the direct MI transmission with no active relay. In particular, even for distances up to 50 m, a data rate above 1 Mbit/s can be achieved. Of course, in a practical system with finite modulation alphabet size and suboptimum filtering the resulting data rate may be somewhat lower. However, such an investigation is beyond the scope of this work.
\section{Conclusion}
\label{sec:5}
In this paper, the potential of an active relaying technique in MI based WUSNs has been investigated, which comprises the traditional relaying of the data by means of amplification or decoding and the passive relaying inherent to MI based transmissions. For this, a suitable system model has been proposed and the system parameters have been optimized in order to maximize the achievable rate. One of our main observations is, that the passively relayed signals have a very little impact on the result due to a much higher path loss. Secondly, a simple AF relaying performs even worse than the direct MI transmission without relay. This is due to a high frequency-selectivity of the path loss, such that the optimal position of the relay in this case is close to the destination. However, then the bottleneck link becomes dominant and the performance degrades. By using a frequency-selective amplification (FF relaying), the signal power can be redistributed in the relay, such that the frequency-selectivity of the channel can be partially compensated. Therefore, this relaying scheme shows a very promising performance, which is close to the upper bound given by the DF relaying. In addition, the possibility of utilizing a full duplex relaying mode with simultaneous signal transmission and reception has been discussed and the performance of this technique has been evaluated.
\bibliographystyle{IEEEtran}
\bibliography{Literatur}

\begin{thebibliography}{10}
\providecommand{\url}[1]{#1}
\csname url@samestyle\endcsname
\providecommand{\newblock}{\relax}
\providecommand{\bibinfo}[2]{#2}
\providecommand{\BIBentrySTDinterwordspacing}{\spaceskip=0pt\relax}
\providecommand{\BIBentryALTinterwordstretchfactor}{4}
\providecommand{\BIBentryALTinterwordspacing}{\spaceskip=\fontdimen2\font plus
\BIBentryALTinterwordstretchfactor\fontdimen3\font minus
  \fontdimen4\font\relax}
\providecommand{\BIBforeignlanguage}[2]{{%
\expandafter\ifx\csname l@#1\endcsname\relax
\typeout{** WARNING: IEEEtran.bst: No hyphenation pattern has been}%
\typeout{** loaded for the language `#1'. Using the pattern for}%
\typeout{** the default language instead.}%
\else
\language=\csname l@#1\endcsname
\fi
#2}}
\providecommand{\BIBdecl}{\relax}
\BIBdecl

\bibitem{WSN_survey}
{I.F. Akyildiz, W. Su, Y. Sankarasubramaniam, and E. Cayirci}, ``{Wireless
  sensor networks: A survey},'' \emph{Comput. Netw. J.}, vol.~38, pp. 393--422,
  March 2002.

\bibitem{WUSN_reschall}
{I.F. Akyildiz and E.P. Stuntebeck}, ``{Wireless underground sensor networks:
  Research challenges},'' \emph{Ad Hoc Netw. J.}, vol.~4, pp. 669--686, July
  2006.

\bibitem{Sig_propag_underground}
{I.F. Akyildiz, Z. Sun, and M.C. Vuran}, ``{Signal propagation techniques for
  wireless underground communication networks},'' \emph{Physical Communication
  Journal (Elsevier)}, vol.~2, pp. 167-- 183, September 2009.

\bibitem{MI_comms_WUSN}
{Z. Sun and I.F. Akyildiz}, ``{Magnetic induction communications for wireless
  underground sensor networks},'' \emph{IEEE Trans. on Antennas and Propag.},
  vol.~58, pp. 2426--2435, July 2010.

\bibitem{cap_perf_near_field}
{H. Jiang and Y. Wang}, ``{Capacity performance of an inductively coupled near
  field communication system},'' in \emph{Proc. of IEEE International Symposium
  of Antenna and Propagation Society}, July 2008.

\bibitem{power_and_cap_MI}
{J.I. Agbinya and M. Mashipour}, ``{Power equations and capacity performance of
  magnetic induction communication systems},'' \emph{Wireless Personal
  Communications Journal}, vol.~64, no.~4, pp. 831--845, 2012.

\bibitem{pract_ch_cap}
{S. Kisseleff, W. Gerstacker, R. Schober, Z. Sun, and I.F. Akyildiz},
  ``{Channel capacity of magnetic induction based wireless underground sensor
  networks under practical constraints},'' in \emph{Proc. of IEEE WCNC 2013},
  April 2013.

\bibitem{modul_MI}
{S. Kisseleff, I.F. Akyildiz, and W. Gerstacker}, ``{On Modulation for Magnetic
  Induction based Transmission in Wireless Underground Sensor Networks},'' in
  \emph{{Proc. of ICC 2014}}, {May} 2014.

\bibitem{MI_waveguide_first}
{E. Shamonina, V. A. Kalinin, K. H. Ringhofer, and L. Solymar},
  ``{Magneto-inductive waveguide},'' \emph{Electronic Letters}, vol.~38, no.~8,
  pp. 371--373, 2002.

\bibitem{agbinya_masihpour_relaying}
{M. Masihpour and J.I. Agbinya}, ``{Cooperative relay in Near Field Magnetic
  Induction: A new technology for embedded medical communication systems},'' in
  \emph{{Proc. of IB2Com}}, December 2010, pp. 1--6.

\bibitem{Interference_polariz}
{S. Kisseleff, I.F. Akyildiz, and W. Gerstacker}, ``{Interference Polarization
  in Magnetic Induction based Wireless Underground Sensor Networks},'' in
  \emph{Proc. of IEEE PIMRC 2013 (SENSA Workshop)}, September 2013.

\bibitem{cooperation_book1}
{F.H.P. Fitzek and M.D. Katz}, \emph{{Cooperation in Wireless Networks:
  Principles and Applications}}.\hskip 1em plus 0.5em minus 0.4em\relax
  {Springer Netherlands}, 2006.

\bibitem{cooperation_book2}
{Y.W.P. Hong, W.J. Huang, and C.C.J. Kuo}, \emph{{Cooperative Communications
  and Networking: Technologies and System Design}}.\hskip 1em plus 0.5em minus
  0.4em\relax {Springer}, 2010.

\bibitem{ff_optimize}
{I. Hammerstrom and A. Wittneben}, ``{On the Optimal Power Allocation for
  Nonregenerative OFDM Relay Links},'' in \emph{{Proc. of ICC 2006}}, June
  2006, pp. 4463--4468.

\bibitem{ff_mimo}
{Z. Fang, Y. Hua, and J.C. Koshy}, ``{Joint Source and Relay Optimization for a
  Non-Regenerative MIMO Relay},'' in \emph{{Proc. of IEEE Workshop on Sensor
  Array and Multichannel Processing}}, July 2006, pp. 239--243.

\bibitem{bconvex}
{S. Boyd and L. Vandenberghe}, \emph{{Convex Optimization}}.\hskip 1em plus
  0.5em minus 0.4em\relax Cambridge University Press, 2004.

\bibitem{chest_MI_2014}
{S. Kisseleff, I.F. Akyildiz, and W. Gerstacker}, ``{Transmitter-Side Channel
  Estimation in Magnetic Induction based Communication Systems},'' in
  \emph{Proc. of IEEE BlackSeaCom 2014}, May 2014.

\end{thebibliography}
\end{document}